\documentclass[11pt,twosided]{article}

\usepackage{amsmath}
\usepackage{amsthm}
\usepackage{amsfonts}
\usepackage{amssymb}
\usepackage[dvips]{epsfig}

\theoremstyle{plain}

\theoremstyle{definition} \theoremstyle{definition}

\renewcommand\thefootnote{\fnsymbol{footnote}}

\numberwithin{equation}{section}

\begin{document}

\title
 {\huge {The evolution of the electric field along optical fiber for the
 type-2 and 3 PAFs in Minkowski 3-space   }}

\author{    Nevin Ertu\u{g} G\"{u}rb\"{u}z$^\dag$ and Dae Won Yoon$^\ddag$ }

\date{}
\maketitle

{\footnotesize
\center { \text{  $^\dag$ Department of Mathematics  and Computer Science}\par
  { \text{ Eskisehir Osmangazi  University}}\par
  {\text{ Eskisehir,  T\"{u}rkiye}}\par
  { \texttt{E-mail address : ngurbuz@ogu.edu.tr}}\par

\text{  $^\ddag$ Department of Mathematics Education and RINS}\par
  { \text{ Gyeongsang National University}}\par
  {\text{ Jinju 52828, Republic of Korea}}\par
  { \texttt{E-mail address : dwyoon@gnu.ac.kr}}\par

}}


\renewcommand\leftmark {\centerline{  \rm  The evolution of the electric field }}
\renewcommand\rightmark {\centerline{ \rm  The evolution of the electric field }}

\renewcommand{\thefootnote}{}


\renewcommand{\thefootnote}{\arabic{footnote}}
\setcounter{footnote}{0}

\begin{abstract}
In this paper, we introduce  the type-2 and the type-3 Positional  Adapted Frame(PAF) of spacelike curve and timelike curve in
Minkowski 3-space.  From these PAFs, we study the evolutions of the electric field vectors of the type-2 and type-3 PAFs.
As a result, we also  investigate  the Fermi-Walker parallel   and the  Lorentz force equation  of the electric field vectors for the type-2 and type-3 PAFs in Minkowski 3-space.
\end{abstract}

\maketitle

\section{Introduction}

It is well known that the Frenet frame of a space curve plays an important role in the study of curve and surface theory, and this frame is  the most well-known frame along a space curve.
However, the Frenet frame is undefined wherever the curvature vanishes, such as at points of inflection or along straight sections of the curve.
In order to solve this problem, Bishop \cite{B}  introduced a new frame along a space curve which is more suitable for
applications, which is called Bishop frame or parallel transport frame.
 After then,  many mathematicians have  studying  various alternative methods of frame of a space curve.
For example,  Arbind et al. \cite{ARS} studied   a general 1-dimensional  higher-order theory for tubes and rod in terms of the hybrid frame of a space curve.  In \cite{KK} authors discussed
 hybrid optical magnetic Lorentz flux by using hybrid frame.
 Also, G\"{u}rb\"{u}z et al. \cite{GMM} presented three formulations associated with the modified  nonlinear Schr\"odinger equation with respect to
 the hybrid frame in Minkowski 3-space. Recently, in \cite{OT1} \"{O}zen and Tosun  introduced the Positional  Adapted Frame (PAF)
 as the the another  frame for the trajectories with non-vanishing angular momentum in  Euclidean 3-space.
This frame is  used to investigate the kinematics of moving particles.

On the other hand, Berry's geometric phase \cite{Be}  is related to the time evolution of a space curve.
The geometric phase of linearly polarized light defines its angle of
rotation. The evolution of an electric field vector is connected with the
geometric phase topic. Also this topic have numerous applications in modern
optic. In last years, many mathematicians are studying  the geometric phase and the
evolution of the electric field vector \cite{BAD}-\cite{G6}, \cite{GY}-\cite{KDA1} etc.

In this paper, we discuss PAF in Minkowski 3-space.  The first author  \cite{G7} in the present paper
studied  the type-1 PAF  in Minkowski space and obtained
the evolution of the electric field according to the type-1 PAF.
Therefore, we want to get two other classes of
the evolutions of the electric field vector according to PAF in Minkowski 3-space  as natural extensions of \"{O}zen and Tosun's formulation


\section{ Construction of the type-2 and 3 PAFs }

In this section, we construct the new frames in terms  of the Frenet frame of the non-null curve  in Minkowski 3-space.

The Minkowski 3-space $\mathbb R_1^3$ is a real space $\mathbb R^3$
 with  the indefinite inner product $\langle \cdot~,  \cdot \rangle_L$
defined on each tangent space  by
$$
\langle \mathbf x, \mathbf y \rangle_L = x_1 y_1 + x_2 y_2 - x_3 y_3,
$$
where ${\mathbf x}=(x_1, x_2, x_3)$ and  ${\mathbf y}=(y_1, y_2, y_3)$  are vectors in $\mathbb R^3_1$.

  A nonzero vector $\mathbf x$ in $\mathbb R^3_1$ is said to be
  spacelike,  timelike or  null if  $\langle \mathbf x, \mathbf x \rangle _L>0$,
   $\langle \mathbf x, \mathbf x \rangle _L<0$
  or  $\langle \mathbf x, \mathbf x \rangle_L =0$, respectively.


Let  $\beta :I\rightarrow \mathbb{R}_{1}^{3}$ be a non-null curve parametrized by the arc-length $s$ in
Minkowski 3-space $\mathbb{R}_{1}^{3}$. Derivative formulae for the Frenet
frame $\left\{ \mathbf{T},\mathbf{N},\mathbf{B}\right\} $ are given by
\begin{equation}
\left(
\begin{array}{c}
\mathbf{T}_{s} \\
\mathbf{N}_{s} \\
\mathbf{B}_{s}%
\end{array}%
\right) =\left(
\begin{array}{ccc}
0 & \varepsilon _{2}\kappa & 0 \\
-\varepsilon _{1}\kappa & 0 & \varepsilon _{3}\tau \\
0 & -\varepsilon _{2}\tau & 0%
\end{array}%
\right) \left(
\begin{array}{c}
\mathbf{T} \\
\mathbf{N} \\
\mathbf{B}%
\end{array}%
\right),  \label{2000}
\end{equation}%
where $\left\langle \mathbf{T},\mathbf{T}\right\rangle _{L}=\varepsilon
_{1}, $ $\left\langle \mathbf{N},\mathbf{N}\right\rangle _{L}=\varepsilon
_{2},$ \ $\left\langle \mathbf{B},\mathbf{B}\right\rangle _{L}=\varepsilon
_{3}$ $,\varepsilon _{i}=\pm 1.$
Here $\kappa $  and $\tau $ are the curvature and the torsion of the non-null curve $\beta$.
On the other hand, the Lorentz cross product implies
$$
\mathbf{T\times }_{L}\mathbf{N} \mathbf{=}\varepsilon _{3}\mathbf{B,} \quad  \mathbf{N\times }_{L}\mathbf{B=}\varepsilon _{1}\mathbf{T,}\quad
\mathbf{B\times }_{L}\mathbf{T=}\varepsilon _{2}\mathbf{N}.
$$
Suppose that a point particle moves along the non-null curve $\beta $ with
  the arc-length $s$  and the time $t$ in Minkowski 3-space $\mathbb{R}_{1}^{3}$.
Then the non-null tangent vector $\mathbf{T}$, the velocity vector $\mathbf{v}$
and the linear momentum vector $\mathbf{M}_{l}$ are given by
\[
\mathbf{T}(s)=\frac{d\mathbf{z}}{ds},\quad \mathbf{v}(t)=\frac{ds}{dt}%
\mathbf{T}(s), \quad \mathbf{M}_{l}(t)=m\frac{ds}{dt}\mathbf{T}(s),
\]%
where $m$ is a constant mass and $\mathbf{z}$ is the position vector of the particle as
\begin{eqnarray*}
\mathbf{z} &=&\varepsilon _{1}\left\langle \beta (s),\mathbf{T}%
(s)\right\rangle _{L}\mathbf{T}(s)+\varepsilon _{2}\left\langle \beta (s),%
\mathbf{N}(s)\right\rangle _{L}\mathbf{N}(s) \\
&&+\varepsilon _{3}\left\langle \beta (s),\mathbf{B}(s)\right\rangle _{L}%
\mathbf{B}(s).
\end{eqnarray*}%
The angular momentum $\mathbf{M}_{a}$ of the particle is the Lorentz cross product
of the position vector $\mathbf{z}$ and the linear momentum vector $\mathbf{M}_{l}$
at the time $t$, and it is expressed as
\begin{eqnarray*}
\mathbf{M}_{a} &=&\mathbf{z\times }_{L}\mathbf{M}_{l} \\
&=&-\varepsilon _{2}\varepsilon _{3}m\frac{ds}{dt}\left\langle \beta (s),%
\mathbf{N}(s)\right\rangle _{L}\mathbf{B}(s) \\
&&+\varepsilon _{2}\varepsilon _{3}m\frac{ds}{dt}\left\langle \beta (s),%
\mathbf{B}(s)\right\rangle _{L}\mathbf{N}(s).
\end{eqnarray*}%
Suppose that the normal component of the angular momentum is not zero, and consider
\begin{eqnarray}
-\mathbf{z} &=&\varepsilon _{1}\left\langle -\beta (s),\mathbf{T}%
(s)\right\rangle _{L}\mathbf{T}(s)  \label{1222} \\
&&+\varepsilon _{2}\left\langle -\beta (s),\mathbf{N}(s)\right\rangle _{L}%
\mathbf{N}(s)+\varepsilon _{3}\left\langle -\beta (s),\mathbf{B}%
(s)\right\rangle _{L}\mathbf{B}(s).  \nonumber
\end{eqnarray}%

  Consider  the projections $\mathbf{w}_{1}$
and $\mathbf{w}_{2}$ on ${\rm Span}\left\{ \mathbf{N},\mathbf{B}\right\} $ and ${\rm Span}\left\{
\mathbf{T},\mathbf{N}\right\} $ of the vector $-\mathbf{z}$. Then these vectors  become
\begin{eqnarray*}
\mathbf{w}_{1} &=&\varepsilon _{2}\left\langle -\beta (s),\mathbf{N}%
(s)\right\rangle _{L}\mathbf{N}(s)+\varepsilon _{3}\left\langle -\beta (s),%
\mathbf{B}(s)\right\rangle _{L}\mathbf{B}(s), \\
\mathbf{w}_{2} &=&\varepsilon _{1}\left\langle -\beta (s),\mathbf{T}%
(s)\right\rangle _{L}\mathbf{T}(s)+\varepsilon _{2}\left\langle -\beta (s),%
\mathbf{N}(s)\right\rangle _{L}\mathbf{N}(s),
\end{eqnarray*}%
respectively. It follows that
\begin{equation}
\mathbf{w}_{1}-\mathbf{w}_{2}=\varepsilon _{3}\left\langle -\beta (s),%
\mathbf{B}(s)\right\rangle _{L}\mathbf{B}(s)+\varepsilon _{1}\left\langle
\beta (s),\mathbf{T}(s)\right\rangle _{L}\mathbf{T}(s).  \label{1901}
\end{equation}%
We define a new vector $\mathbf{H}$ as follows:%
\begin{eqnarray*}
\mathbf{H} &=&\frac{\mathbf{w}_{1}-\mathbf{w}_{2}}{\sqrt{\left\vert
\left\langle \mathbf{w}_{1}-\mathbf{w}_{2},\mathbf{w}_{1}-\mathbf{w}%
_{2}\right\rangle _{L}\right\vert }} \\
&=&\varepsilon _{1}\frac{\left\langle \beta (s),\mathbf{T}(s)\right\rangle
_{M}}{\sqrt{\left\vert \varepsilon _{1}\left\langle \beta (s),\mathbf{T}%
(s)\right\rangle _{L}^{2}+\varepsilon _{3}\left\langle \beta (s),\mathbf{B}%
(s)\right\rangle _{L}^{2}\right\vert }}\mathbf{T}(s) \\
&&+\varepsilon _{3}\frac{\left\langle -\beta (s),\mathbf{B}(s)\right\rangle
_{L}}{\sqrt{\left\vert \varepsilon _{1}\left\langle \beta (s),\mathbf{T}%
(s)\right\rangle _{L}^{2}+\varepsilon _{3}\left\langle \beta (s),\mathbf{B}%
(s)\right\rangle _{L}^{2}\right\vert }}\mathbf{B}(s)
\end{eqnarray*}%
and  we take an another vector $\mathbf{D}=\varepsilon _{3}\mathbf{H}\times _{L}\mathbf{N}$ given by
\begin{eqnarray*}
\mathbf{D} &=&\varepsilon _{1}\frac{\left\langle \beta (s),\mathbf{T}%
(s)\right\rangle _{L}}{\sqrt{\left\vert \varepsilon _{1}\left\langle \beta
(s),\mathbf{T}(s)\right\rangle _{L}^{2}+\varepsilon _{3}\left\langle \beta
(s),\mathbf{B}(s)\right\rangle _{L}^{2}\right\vert }}\mathbf{B}(s) \\
&&+\varepsilon _{1}\frac{\left\langle \beta (s),\mathbf{B}(s)\right\rangle
_{L}}{\sqrt{\left\vert \varepsilon _{1}\left\langle \beta (s),\mathbf{T}%
(s)\right\rangle _{L}^{2}+\varepsilon _{3}\left\langle \beta (s),\mathbf{B}%
(s)\right\rangle _{L}^{2}\right\vert }}\mathbf{T}(s).
\end{eqnarray*}%
In this case the moving frame $\{ {\mathbf H}, {\mathbf N}, {\mathbf D}\}$ along the non-null curve $\beta$ is called the type-2 Positional  Adapted Frame (type-2 PAF) in Minkowski 3-space.

Now, we give the relationship between the Frenet frame and the type-2 PAF frame.

First of all, if $\mathbf{N}$ is the  timelike vector and  $\mathbf{H}$, $\mathbf{D}$ are the
spacelike vectors ($\mathbf{T}$ and $\mathbf{B}$ are the spacelike
vectors), then it can be written by
\[
\left(
\begin{array}{c}
\mathbf{H} \\
\mathbf{N} \\
\mathbf{D}%
\end{array}%
\right) =\left(
\begin{array}{ccc}
\cos \phi & 0 & -\sin \phi \\
0 & 1 & 0 \\
\sin \phi & 0 & \cos \phi%
\end{array}%
\right) \left(
\begin{array}{c}
\mathbf{T} \\
\mathbf{N} \\
\mathbf{B}%
\end{array}%
\right),
\]%
where $\phi$ is an angle between $\mathbf D$ and $\mathbf B$.
On the other hand, the type-2 PAF frame apparatus $p_{1},$ $p_{2},$ $p_{3}$ are given by
\begin{eqnarray*}
p_{1} &=&\kappa (s)\cos \phi +\tau (s)\sin \phi \\
p_{2} &=&-\phi ^{\prime } \\
p_{3} &=&-\kappa (s)\sin \phi +\tau (s)\cos \phi .
\end{eqnarray*}
Secondly, if $\mathbf{D}$ is the timelike vector and  $\mathbf{H}$, $\mathbf{N}$ are the
spacelike vectors, $(\mathbf{B}$ is the timelike vector), then we have the relationship as follows:
\[
\left(
\begin{array}{c}
\mathbf{H} \\
\mathbf{N} \\
\mathbf{D}%
\end{array}%
\right) =\left(
\begin{array}{ccc}
\cosh \phi & 0 & \sinh \phi \\
0 & 1 & 0 \\
\sinh \phi & 0 & \cosh \phi%
\end{array}%
\right) \left(
\begin{array}{c}
\mathbf{T} \\
\mathbf{N} \\
\mathbf{B}%
\end{array}%
\right),
\]%
where $\phi $ is the angle between the vectors $\mathbf{D}$ and $\mathbf{B}.$
The type-2 PAF frame apparatus $p_{1},$ $p_{2},$ $p_{3}$ are given by
\begin{eqnarray*}
p_{1} &=&\kappa (s)\cosh \phi (s)-\tau (s)\sinh \phi (s) \\
p_{2} &=&-\phi ^{\prime }(s) \\
p_{3} &=&-\kappa (s)\sinh \phi (s)+\tau (s)\cosh \phi (s).
\end{eqnarray*}%
Also, the derivative formulas of the type-2 PAF frame $\left\{ \mathbf{H},\mathbf{N%
},\mathbf{D}\right\} $ of the non-null curve $\beta$  in  Minkowski 3-space are
expressed as
\begin{equation}
\left(
\begin{array}{c}
\mathbf{H}_{s} \\
\mathbf{N}_{s} \\
\mathbf{D}_{s}%
\end{array}%
\right) =\left(
\begin{array}{ccc}
0 & \varepsilon _{2}p_{1} & \varepsilon _{3}p_{2} \\
-\varepsilon _{1}p_{1} & 0 & \varepsilon _{3}p_{3} \\
-\varepsilon _{1}p_{2} & -\varepsilon _{2}p_{3} & 0%
\end{array}%
\right) \left(
\begin{array}{c}
\mathbf{H} \\
\mathbf{N} \\
\mathbf{D}%
\end{array}%
\right),  \label{1*}
\end{equation}
where
\begin{eqnarray*}
\left\langle \mathbf{H},\mathbf{H}\right\rangle _{L} &=&\varepsilon _{1},\quad
\left\langle \mathbf{N},\mathbf{N}\right\rangle _{L}=\varepsilon
_{2},\quad \left\langle \mathbf{D},\mathbf{D}\right\rangle
_{L}=\varepsilon _{3}, \\
\left\langle \mathbf{H},\mathbf{N}\right\rangle _{L} &=&\left\langle \mathbf{%
N},\mathbf{D}\right\rangle _{L}=\left\langle \mathbf{H},\mathbf{D}%
\right\rangle _{L}=0.
\end{eqnarray*}%
As a similar method, we can define the new frame in terms of the Frenet frame.

Suppose that the binormal component of the angular momentum is not zero. The
projections $\mathbf{\Gamma } _{1}$ and $\mathbf{\Gamma }_{2}$ on ${\rm Span}\left\{ \mathbf{N},\mathbf{B}\right\} $ and ${\rm Span}\left\{
\mathbf{T},\mathbf{B}\right\} $ of the vector $-\mathbf{z}$  are  given by, respectively
\begin{eqnarray*}
\mathbf{\Gamma }_{1} &=&\varepsilon _{2}\left\langle -\beta (s),\mathbf{N}%
(s)\right\rangle _{L}\mathbf{N}(s)+\varepsilon _{3}\left\langle -\beta (s),%
\mathbf{B}(s)\right\rangle _{L}\mathbf{B}(s). \\
\mathbf{\Gamma }_{2} &=&\varepsilon _{1}\left\langle -\beta (s),\mathbf{T}%
(s)\right\rangle _{L}\mathbf{T}(s)+\varepsilon _{3}\left\langle -\beta (s),%
\mathbf{N}(s)\right\rangle _{L}\mathbf{B}(s).
\end{eqnarray*}%
From this,
\begin{equation}
\mathbf{\Gamma }_{1}-\mathbf{\Gamma }_{2}=\varepsilon _{2}\left\langle
-\beta (s),\mathbf{N}(s)\right\rangle _{L}\mathbf{N}(s)+\varepsilon
_{1}\left\langle \beta (s),\mathbf{T}(s)\right\rangle _{L}\mathbf{T}(s).
\end{equation}%
it follows that  we define the new vector $\mathbf{F}$  as follows:%
\begin{eqnarray*}
\mathbf{F} &=&\frac{\mathbf{\Gamma }_{1}-\mathbf{\Gamma }_{2}}{\sqrt{%
\left\vert \left\langle \mathbf{\Gamma }_{1}-\mathbf{\Gamma }_{2},\mathbf{%
\Gamma }_{1}-\mathbf{\Gamma }_{2}\right\rangle _{L}\right\vert }} \\
&=&\varepsilon _{1}\frac{\left\langle \beta (s),\mathbf{T}(s)\right\rangle
_{L}}{\sqrt{\left\vert \varepsilon _{1}\left\langle \beta (s),\mathbf{T}%
(s)\right\rangle _{L}^{2}+\varepsilon _{2}\left\langle \beta (s),\mathbf{N}%
(s)\right\rangle _{L}^{2}\right\vert }}\mathbf{T}(s) \\
&&+\varepsilon _{2}\frac{\left\langle -\beta (s),\mathbf{N}(s)\right\rangle
_{L}}{\sqrt{\left\vert \varepsilon _{1}\left\langle \beta (s),\mathbf{T}%
(s)\right\rangle _{L}^{2}+\varepsilon _{3}\left\langle \beta (s),\mathbf{N}%
(s)\right\rangle _{L}^{2}\right\vert }}\mathbf{N}(s).
\end{eqnarray*}%
Also, the binormal vector $\mathbf B$ and the vector $\mathbf H$ lead to  the another vector $\mathbf{P}$:
\begin{eqnarray*}
\mathbf{P} &=&\varepsilon _{1}\mathbf{F}\times _{L}\mathbf{B}\\
&=&-\varepsilon _{2}\frac{\left\langle \beta (s),\mathbf{T}%
(s)\right\rangle _{L}}{\sqrt{\left\vert \varepsilon _{1}\left\langle \beta
(s),\mathbf{T}(s)\right\rangle _{L}^{2}+\varepsilon _{3}\left\langle \beta
(s),\mathbf{B}(s)\right\rangle _{L}^{2}\right\vert }}\mathbf{N}(s) \\
&&-\varepsilon _{2}\frac{\left\langle \beta (s),\mathbf{N}(s)\right\rangle
_{M}}{\sqrt{\left\vert \varepsilon _{1}\left\langle \beta (s),\mathbf{T}%
(s)\right\rangle _{L}^{2}+\varepsilon _{3}\left\langle \beta (s),\mathbf{B}%
(s)\right\rangle _{L}^{2}\right\vert }}\mathbf{T}(s).
\end{eqnarray*}%
In this case, the moving frame $\{ {\mathbf P}, {\mathbf F}, {\mathbf B}\}$ along the non-null curve $\beta$ is called the type-3 Positional  Adapted Frame (type-3 PAF) in Minkowski 3-space.


If $\mathbf{B}$ is the timelike vector and $\mathbf{P}$, $\mathbf{F}$ are the
spacelike vectors, then the relationship between the Frenet frame and the type-3 PAF frame  are expressed by
\[
\left(
\begin{array}{c}
\mathbf{P} \\
\mathbf{F} \\
\mathbf{B}%
\end{array}%
\right) =\left(
\begin{array}{ccc}
\cos \phi & -\sin \phi & 0 \\
\sin \phi & \cos \phi & 0 \\
0 & 0 & 1%
\end{array}%
\right) \left(
\begin{array}{c}
\mathbf{T} \\
\mathbf{N} \\
\mathbf{B}%
\end{array}%
\right),
\]%
where $\phi $ is the angle between the vectors $\mathbf{F}$ and $\mathbf{N}.$
Also, the type-3 PAF frame apparatus $n_{1},$ $n_{2},$ $n_{3}$ are given by
\begin{eqnarray*}
n_{1} &=&\kappa -\phi ^{\prime } \\
n_{2} &=&-\tau \sin \phi \\
n_{3} &=&\tau \cos \phi.
\end{eqnarray*}%
If $\mathbf{F}$ is the timelike vector ($\mathbf{N}$ is the timelike) and  $\mathbf{P}$,
$\mathbf{B}$ are the spacelike vectors, then we have
\[
\left(
\begin{array}{c}
\mathbf{P} \\
\mathbf{F} \\
\mathbf{B}%
\end{array}%
\right) =\left(
\begin{array}{ccc}
\cosh \phi & \sinh \phi & 0 \\
\sinh \phi & \cosh \phi & 0 \\
0 & 0 & 1%
\end{array}%
\right) \left(
\begin{array}{c}
\mathbf{T} \\
\mathbf{N} \\
\mathbf{B}%
\end{array}%
\right),
\]%
it follows that the type-3 PAF frame apparatus $n_{1},$ $n_{2},$ $n_{3}$ are given by
\begin{eqnarray*}
n_{1} &=&\kappa -\phi ^{\prime } \\
n_{2} &=&\tau \sinh \phi \\
n_{3} &=&\tau \cosh \phi.
\end{eqnarray*}%
On the other hand, the derivative formulas of the type-3 PAF frame $\left\{ \mathbf{P},\mathbf{F%
},\mathbf{B}\right\} $ of the non-null curve $\beta$ in  Minkowski 3-space become
\[
\left(
\begin{array}{c}
\mathbf{P}_{s} \\
\mathbf{F}_{s} \\
\mathbf{B}_{s}%
\end{array}%
\right) =\left(
\begin{array}{ccc}
0 & \varepsilon _{2}n_{1} & \varepsilon _{3}n_{2} \\
-\varepsilon _{1}n_{1} & 0 & \varepsilon _{3}n_{3} \\
-\varepsilon _{1}n_{2} & -\varepsilon _{2}n_{3} & 0%
\end{array}%
\right) \left(
\begin{array}{c}
\mathbf{P} \\
\mathbf{F} \\
\mathbf{B}%
\end{array}%
\right),
\]
where
\begin{eqnarray*}
\left\langle \mathbf{P},\mathbf{P}\right\rangle _{L} &=&\varepsilon _{1},\quad
\left\langle \mathbf{F},\mathbf{F}\right\rangle _{L}=\varepsilon
_{2},\quad \left\langle \mathbf{B},\mathbf{B}\right\rangle
_{L}=\varepsilon _{3}, \\
\left\langle \mathbf{P},\mathbf{F}\right\rangle _{L} &=&\left\langle \mathbf{%
F},\mathbf{B}\right\rangle _{L}=\left\langle \mathbf{P},\mathbf{B}%
\right\rangle _{L}=0.
\end{eqnarray*}

\section{ The evolution of electric field for the type-2 PAF}

In this section, we study   the evolution of the electric field vector with respect to the type-2 PAF  in Minkowski 3-space.
To get results, we split it into three cases according to the type-2 PAF.




\vskip 0.3 cm

\textbf{Case I.} Consider an optical fiber $\mathcal{O}$  described by the spacelike curve $\beta $ and
the timelike binormal vector of  the type-2 PAF  in  Minkowski 3-space $\mathbb{R}_{1}^{3}$.

Suppose that the electric field vector $\mathbf{E}^{(2.Paf)}$ of  the type-2 PAF is
perpendicular to the spacelike vector $\mathbf{H}$ with the timelike vector
$\mathbf{D}$, that is,
\begin{equation}
\left\langle \mathbf{E}^{(2.Paf)}\mathbf{,H}\right\rangle _{L}=0.
\label{79*3}
\end{equation}%
On the other hand, the general evolution of the electric field vector $\mathbf{E}^{(2.Paf)}$ for
the type-2 PAF  in $\mathbb{R}_{1}^{3}$ is expressed by
\begin{equation}
\mathbf{E}_{s}^{(2.Paf)}=a_{1}\mathbf{H}+a_{2}\mathbf{N}+a_{3}\mathbf{D},
\label{2.1*3}
\end{equation}%
where $a_{1},$ $a_{2}$ and $a_{3}$ are the arbitrary
smooth functions. Consider that no various loss mechanism along  the
optic fiber for   the electric field vector $\mathbf{E}^{(2.Paf)}$ of the type-2 PAF in  Minkowski
3-space, then it can be written by

\begin{equation}
\left\langle \mathbf{E}^{(2.Paf)},\mathbf{E}^{(2.Paf)}\right\rangle
_{L}={\rm constant.}  \label{69690*3}
\end{equation}%
Using Eq.(\ref{79*3}) and Eq.(\ref{2.1*3}), we obtain
\begin{equation}
a_{1}=-p_{1}\left\langle \mathbf{E}^{(2.Paf)},\mathbf{N}\right\rangle
_{L}+p_{2}\left\langle \mathbf{E}^{(2.Paf)},\mathbf{D}\right\rangle _{L},
\label{1233*}
\end{equation}%
where $\left\langle \mathbf{E}^{(2.Paf)},\mathbf{N}\right\rangle _{L}\neq 0$ and
 $\left\langle \mathbf{E}^{(2.Paf)},\mathbf{D}\right\rangle _{L}\neq 0$.
Taking the derivative  with respect to  $s$ of  Eq.(\ref{69690*3}) and using Eq.(\ref%
{2.1*3}), we also have
\begin{equation}
a_{2}=\sigma \left\langle \mathbf{E}^{(2.Paf)},\mathbf{D}\right\rangle _{L}, \quad
a_{3}=-\sigma \left\langle \mathbf{E}^{(2.Paf)},\mathbf{N}%
\right\rangle _{L},  \label{50*138}
\end{equation}%
where $\sigma $ is a parameter. When Eqs.(\ref{1233*}) and (\ref{50*138})
are substituted in Eq.(\ref{2.1*3}), the evolution of the electric field
vector $\mathbf{E}^{(2.Paf)}$ for the type-2 PAF  is expressed as
\begin{eqnarray}
\mathbf{E}_{s}^{(2.Paf)} &=&\left( -p_{1}\left\langle \mathbf{E}^{(2.Paf)},%
\mathbf{N}\right\rangle _{L}+p_{2}\left\langle \mathbf{E}^{(2.Paf)},\mathbf{D%
}\right\rangle _{L}\right) \mathbf{H}  \label{345*7} \\
&&+\sigma \left\langle \mathbf{E}^{(2.Paf)},\mathbf{D}\right\rangle _{L}%
\mathbf{N}-\sigma \left\langle \mathbf{E}^{(2.Paf)},\mathbf{N}\right\rangle
_{L}\mathbf{D}.  \nonumber
\end{eqnarray}%
On the other hand, the Fermi-Walker derivative $^{FW}\mathbf{E}_{s}^{(2.Paf)}$ of  the electric field $\mathbf{E}%
^{(2.Paf)}$ with respect to the type-2 PAF  in $\mathbb{R}_{1}^{3}$ is
given by
\begin{equation}
^{FW}\mathbf{E}_{s}^{(2.Paf)}=\mathbf{E}_{s}^{(2.Paf)}-\left\langle \mathbf{H%
},\mathbf{E}^{(2.Paf)}\right\rangle _{L}\mathbf{H}_{s}+\left\langle \mathbf{H%
}_{s},\mathbf{E}^{(2.Paf)}\right\rangle _{L}\mathbf{H}.  \label{1251}
\end{equation}%
The electric field $\mathbf{E}^{(2.Paf)}$ is
Fermi-Walker parallel of the type-2 PAF if and only if  $^{FW}\mathbf{E}_{s}^{(2.Paf)}=0.$
If the electric field $\mathbf{E}^{(2.Paf)}$ is
Fermi-Walker parallel, then   Eq.(\ref%
{79*3}) and Eq.(\ref{1251}) imply
\begin{equation}
\mathbf{E}_{s}^{(2.Paf)}=-\left\langle \mathbf{H}_{s},\mathbf{E}%
^{(2.Paf)}\right\rangle _{L}\mathbf{H.}  \label{68681}
\end{equation}%
Using Eq.(\ref{79*3}) the  electric field $\mathbf{E}^{(2.Paf)}$ of the type-2 PAF
is expressed by
\begin{equation}
\mathbf{E}^{(2.Paf)}=\mathbf{E}^{(2.Paf)N}\mathbf{N}-\mathbf{E}^{(2.Paf)D}%
\mathbf{D},  \label{789*1}
\end{equation}%
where
\[
E^{(2.Paf)N}=\left\langle \mathbf{E}^{(2.Paf)},\mathbf{N}\right\rangle _{L}, \quad
E^{(2.Paf)D}=\left\langle \mathbf{E}^{(2.Paf)},\mathbf{D}%
\right\rangle _{L}.
\]%

\noindent Taking derivative of Eq. (\ref {789*1})  with respect to $s$, the variation of the electric field vector $\mathbf{E}^{(2.Paf)}$
for the first case is given  by
\begin{eqnarray}
\mathbf{E}_{s}^{(2.Paf)} &\mathfrak{=}&\mathbf{(}%
E_{s}^{(2.Paf)N}+p_{3}E^{(2.Paf)D})\mathbf{N}  \label{6751} \\
&&-\mathbf{(}p_{3}E^{(2.Paf)N}+E_{s}^{(2.Paf)D})\mathbf{D}  \nonumber \\
&&+\mathbf{(}p_{1}E^{(2.Paf)N}+p_{2}E^{(2.Paf)D})\mathbf{H}.  \nonumber
\end{eqnarray}%
Via Eqs.(\ref{68681}) and (\ref{6751}), one finds
\begin{equation}
E_{s}^{(2.Paf)N}=-p_{3}E^{(2.Paf)D},\quad
E_{s}^{(2.Paf)D}=-p_{3}E^{(2.Paf)N},  \label{4566}
\end{equation}%
it follows that  Eq.(\ref{4566}) gives a rotation of the polarization plane by an angle $%
p_{3} $ for the first case with the type-2 PAF  in $\mathbb{R}_{1}^{3}.$

A magnetic field vector is given by a closed 2-form $\mathcal{C}$ in 3-dimensional (pseudo-)Riemannian manifold $M$.
 The Lorentz force of a magnetic field $%
\mathbf{V}$ is described by skew-symmetric operator $\Phi $, that is,
\[
\left\langle \Phi (\mathbf{X}),\mathbf{Y}\right\rangle =\mathcal{C}(\mathbf{X%
},\mathbf{Y})
\]%
for all $\mathbf{X},$ $\mathbf{Y\in }\chi (M)$   and $\Phi (\mathbf{X})=\mathbf{V}%
\times \mathbf{X}.$

The Lorentz force equation $\Phi \mathbf{E}^{(2.Paf)}$ of the electric field
vector $\mathbf{E}^{(2.Paf)}$ with the magnetic vector field $\mathbf{V}%
^{(1)}$ of the type-2 PAF in Minkowski 3-space  is
\begin{equation}
\Phi ^{H}\mathbf{E}^{(2.Paf)}=\mathbf{E}_{s}^{(2.Paf)}=\mathbf{V}%
^{(1)}\times _{L}\mathbf{E}^{(2.Paf)}.  \label{123*6}
\end{equation}%
The curve $E^{(2.Paf)}V^{(1)}$ moved along the electromagnetic trajectory is
called the  electromagnetic curve with  the type-2 PAF
in $\mathbb{R}_{1}^{3}$.
The Lorentz force equations $\Phi ^H$
along  the optical fiber via Eqs.(\ref{345*7}) and (\ref{123*6})   with the type-2 PAF  is obtained by
\begin{equation}
\left[
\begin{array}{c}
\Phi ^{H}(\mathbf{H)} \\
\Phi ^{H}(\mathbf{N}) \\
\Phi ^{H}(\mathbf{D})%
\end{array}%
\right] =\left[
\begin{array}{ccc}
0 & p_{1} & -p_{2} \\
-p_{1} & 0 & -\sigma \\
-p_{2} & -\sigma & 0%
\end{array}%
\right] \left[
\begin{array}{c}
\mathbf{H} \\
\mathbf{N} \\
\mathbf{D}%
\end{array}%
\right]  \label{7975}
\end{equation}%
and the magnetic vector field $\mathbf{V}^{(1)}$ via Eq.(\ref{7975}) with
respect to the type-2 PAF  becomes
\[
\mathbf{V}^{(1)}=\sigma \mathbf{H-}p_{2}\mathbf{N}-p\mathbf{_{1}D}.
\]

\vskip 0.3 cm




\textbf{Case II.} Suppose that an optical fiber $\mathcal{O}$ can be described by a spacelike curve $\beta $
with the timelike normal vector  for the type-2 PAF  in  Minkowski 3-space $%
\mathbb{R}_{1}^{3}$.

Now, we consider the electric field vector $\mathbf{E}^{(2.Paf)}$
perpendicular to the timelike vector $\mathbf{N}$ according to the type-
2 PAF,  that is,
\begin{equation}
\left\langle \mathbf{E}^{(2.Paf)}\mathbf{,N}\right\rangle _{L}=0.
\label{79*31}
\end{equation}%
The general variation of the electric field vector $\mathbf{E}^{(2.Paf)}$
for the type-2 PAF frame in $\mathbb{R}_{1}^{3}$ is given by
\begin{equation}
\mathbf{E}_{s}^{(2.Paf)}=b_{1}\mathbf{H}+b_{2}\mathbf{N}+b_{3}\mathbf{D},
\label{2.1*31}
\end{equation}%
where $b_{1},$ $b_{2}$ and $b_{3}$ are the arbitrary
smooth functions. Consider  no various loss mechanism along with the
optical fiber for the type-2 PAF  with $\mathbf{E}^{(2.Paf)}\perp \mathbf{N%
}$. Then from Eqs. (\ref{69690*3}), (\ref{79*31}) and (\ref{2.1*31}) we have
\begin{equation}
b_{2}=-p_{1}\left\langle \mathbf{E}^{(2.Paf)},\mathbf{H}\right\rangle
_{L}+p_{3}\left\langle \mathbf{E}^{(2.Paf)},\mathbf{D}\right\rangle _{L}
\label{1233**1}
\end{equation}%
with $\left\langle \mathbf{E}^{(2.Paf)},\mathbf{H}\right\rangle _{M}\neq 0$ and
 $\left\langle \mathbf{E}^{(2.Paf)},\mathbf{D}\right\rangle _{M}\neq 0$.
By differentiating Eq.(\ref{69690*3}) with respect to  $s$  and  using Eq.(\ref
{2.1*31}) we also obtain
\begin{equation}
b_{1}=\rho \left\langle \mathbf{E}^{(2.Paf)},\mathbf{D}\right\rangle _{L},\quad
b_{3}=\rho \left\langle \mathbf{E}^{(2.Paf)},\mathbf{H}%
\right\rangle _{L},  \label{50*13}
\end{equation}%
where $\rho $ is a parameter.
If Eqs. (\ref{1233**1}) and (\ref{50*13}) are
substituted in Eq.(\ref{2.1*31}), thus  the evolution of the electric field $
\mathbf{E}^{(2.Paf)}$  for the type-2 PAF
 in $\mathbb{R}_{1}^{3}$ is derived by
\begin{eqnarray}
\mathbf{E}_{s}^{(2.Paf)} &=&\rho \left\langle \mathbf{E}^{(2.Paf)},\mathbf{D}
\right\rangle _{L}\mathbf{H}  \label{345*71} \\
&&+\left(p_{3}\left\langle \mathbf{E}^{(2.Paf)},\mathbf{D}\right\rangle
_{L}-p_{1}\left\langle \mathbf{E}^{(2.Paf)},\mathbf{H}\right\rangle _{L} \right)
\mathbf{N}  \nonumber \\
&&+\rho \left\langle \mathbf{E}^{(2.Paf)},\mathbf{H}\right\rangle _{L}%
\mathbf{D}.  \nonumber
\end{eqnarray}%
The Fermi-Walker derivative of the electric field $\mathbf{E}^{(2.Paf)}$
with respect to the type-2 PAF  for the second case in $\mathbb{R}%
_{1}^{3}$ is given by

\begin{eqnarray}
^{FW}\mathbf{E}_{s}^{(2.Paf)} &=&\mathbf{E}_{s}^{(2.Paf)}+\left\langle
\mathbf{N},\mathbf{E}^{(2.Paf)}\right\rangle _{L}\mathbf{N}_{s}
\label{12511} \\
&&-\left\langle \mathbf{N}_{s},\mathbf{E}^{(2.Paf)}\right\rangle _{L}\mathbf{%
N}.  \nonumber
\end{eqnarray}%
If the electric field $\mathbf{E}^{(2.Paf)}$ of the type-2 PAF is the Fermi-Walker
parallel for second case,  Eq.(\ref {79*31}) and Eq.(\ref{12511}) imply
\begin{eqnarray}
\mathbf{E}_{s}^{(2.Paf)} &=&\left\langle \mathbf{N}_{s},\mathbf{E}%
^{(2.Paf)}\right\rangle _{L}\mathbf{N}  \label{686811} \\
&=&-p_{1}\left\langle \mathbf{E}^{(2.Paf)},\mathbf{H}\right\rangle
_{L}+p_{3}\left\langle \mathbf{E}^{(2.Paf)},\mathbf{D}\right\rangle _{L}.
\nonumber
\end{eqnarray}%
Furthermore, by  Eq.(\ref{79*31}) the electric field $\mathbf{E}^{(2.Paf)}$ of the type-2 PAF
for the second case leads to
\begin{equation}
\mathbf{E}^{(2.Paf)}=\mathbf{E}^{(2.Paf) H}\mathbf{H}+\mathbf{E}^{(2.Paf)D}%
\mathbf{D},  \label{789*11}
\end{equation}%
where
\[
E^{(2.Paf)H}=\left\langle \mathbf{E}^{(2.Paf)},\mathbf{H}\right\rangle _{L}, \quad
E^{(2.Paf)D}=\left\langle \mathbf{E}^{(2.Paf)},\mathbf{D}%
\right\rangle _{L}.
\]%
Taking derivative of Eq.(\ref{789*11})  with respect to  $s$, the
change of the electric field vector $\mathbf{E}^{(2.Paf)}$ in  Minkowski 3-space can be
expressed as
\begin{eqnarray}
\mathbf{E}_{s}^{(2.Paf)} &\mathfrak{=}&\mathbf{(}%
E_{s}^{(2.Paf)H}-p_{2}E^{(2.Paf)D})\mathbf{H}  \label{67511} \\
&&+\mathbf{(}p_{3}E^{(2.Paf)D}-p_{1}E^{(2.Paf)H})\mathbf{N}  \nonumber \\
&&+\mathbf{(}p_{2}E^{(2.Paf)H}+E_{s}^{(2.Paf)D})\mathbf{D},  \nonumber
\end{eqnarray}%
it follows that 
from Eq.(\ref{686811}) we obtain
\begin{equation}
E_{s}^{(2.Paf)H}=p_{2}E^{(2.Paf)D},\quad
E_{s}^{(2.Paf)D}=-p_{2}E^{(2.Paf)H}.  \label{45661}
\end{equation}%
Therefore, Eq.(\ref{45661}) gives  a rotation of the polarization plane by an angle $%
p_{2}$ for the second class with the type-2 PAF.

On the other hand, the Lorentz force equation $\Phi \mathbf{E}^{(2.Paf)}$ of the electric field
vector $\mathbf{E}^{(2.Paf)}$    of  the  magnetic
vector field $\mathbf{V}^{(2)}$ for the second case is
\begin{equation}
\Phi ^{N}\mathbf{E}^{(2.Paf)}=\mathbf{E}_{s}^{(2.Paf)}=\mathbf{V}%
^{(2)}\times _{L}\mathbf{E}^{(2.Paf)}.  \label{123*61}
\end{equation}%
The curve $E^{(2.Paf)}V^{(2)}$ travelled along the electromagnetic
trajectory is described by the  electromagnetic curve according
to the type-2 PAF  in $\mathbb{R}_{1}^{3}$. The  Lorentz
force equations $\Phi ^{N}$  of  the type-2 PAF along  the optical fiber via Eqs.(\ref%
{345*71}) and (\ref{123*61}) with the second case  are
given by

\begin{equation}
\left[
\begin{array}{c}
\Phi ^{N}(\mathbf{H)} \\
\Phi ^{N}(\mathbf{N}) \\
\Phi ^{N}(\mathbf{D})%
\end{array}%
\right] =\left[
\begin{array}{ccc}
0 & p_{1} & -\rho \\
p_{1} & 0 & -p_{3} \\
-\rho & -p_{3} & 0%
\end{array}%
\right] \left[
\begin{array}{c}
\mathbf{H} \\
\mathbf{N} \\
\mathbf{D}%
\end{array}%
\right].  \label{79751}
\end{equation}%
Also, using Eq.(\ref{79751}) the magnetic vector field $\mathbf{V}^{(2)}$   with respect to the type-2 PAF for the
second case is obtained by
\[
\mathbf{V}^{(2)}=\mathbf{-}p_{3}\mathbf{H+}\rho \mathbf{N}-p\mathbf{_{1}D}.
\]




\vskip 0.3cm

\textbf{Case III.} Let   an optical fiber $\mathcal{O}$ can be described by a spacelike curve $\beta $
with the timelike binormal vector with respect to the type-2 PAF  in the
Minkowski 3-space $\mathbb{R}_{1}^{3}$.

We assume that  the electric field vector $\mathbf{E}%
^{(2.Paf)}$ is perpendicular to the timelike vector $\mathbf{D}$, that is,
\begin{equation}
\left\langle \mathbf{E}^{(2.Paf)}\mathbf{,D}\right\rangle _{L}=0,
\label{1010}
\end{equation}%
and  we also consider   no various loss mechanism along the optical fiber  of the electric field vector $\mathbf{E}$ of the type-2 PAF with the third case in  Minkowski
3-space. Then one finds
\begin{equation}
\left\langle \mathbf{E}^{(2.Paf)}\mathbf{,E}^{(2.Paf)}\right\rangle _{L}=0.
\label{1212}
\end{equation}%
The general evolution of the electric field vector $\mathbf{E}^{(2.Paf)}$ of the type-2 PAF  in $\mathbb{R}_{1}^{3}$ is given by
\begin{equation}
\mathbf{E}_{s}^{(2.Paf)}=c_{1}\mathbf{H}+c_{2}\mathbf{N}+c_{3}\mathbf{D},
\label{1011}
\end{equation}%
where $c_{1},$ $c_{2}$ and $c_{3}$ are the arbitrary smooth
functions.
Using Eqs.(\ref{1010}), (\ref{1212}) and (\ref{1011}), we get
\begin{eqnarray}
c_{1} &=&\xi \left\langle \mathbf{E}^{(2.Paf)},\mathbf{N}\right\rangle _{L}, \quad
c_{2}=-\xi \left\langle \mathbf{E}^{(2.Paf)},\mathbf{H}%
\right\rangle _{L},  \label{1233**} \\
c_{3} &=&-p_{2}\left\langle \mathbf{E}^{(2.Paf)},\mathbf{H}\right\rangle
_{L}-p_{3}\left\langle \mathbf{E}^{(2.Paf)},\mathbf{N}\right\rangle _{L}
\label{1234}
\end{eqnarray}%
with  $\left\langle \mathbf{E}^{(2.Paf)},\mathbf{H}\right\rangle _{L}\neq 0$ and
 $\left\langle \mathbf{E}^{(2.Paf)},\mathbf{N}\right\rangle _{L}\neq 0$. Here
$\xi $ is a parameter.
When Eqs.(\ref{1233**}) and (\ref{1234}) are written
in Eq.(\ref{1011}), the evolution of the electric field vector $\mathbf{E}%
^{(2.Paf)}$ for the third case with respect to the type-2 PAF  in
$\mathbb{R}_{1}^{3}$ is found by
\begin{eqnarray}
\mathbf{E}_{s}^{(2.Paf)} &=&\xi \left\langle \mathbf{E}^{(2.Paf)},\mathbf{N}%
\right\rangle _{L}\mathbf{H}-\xi \left\langle \mathbf{E}^{(2.Paf)},\mathbf{H}%
\right\rangle _{L}\mathbf{N}  \label{345*78} \\
&&-\left(p_{3}\left\langle \mathbf{E}^{(2.Paf)},\mathbf{N}\right\rangle
_{L}+p_{2}\left\langle \mathbf{E}^{(2.Paf)},\mathbf{H}\right\rangle _{L}\right)
\mathbf{D}.  \nonumber
\end{eqnarray}%
The Fermi-Walker derivative of the electric field $\mathbf{E}%
^{(2.Paf)}$ with respect to the type-2 PAF  for third case in $\mathbb{R%
}_{1}^{3}$ is given by
\begin{eqnarray}
^{FW}\mathbf{E}_{s}^{(2.Paf)} &=&\mathbf{E}_{s}^{(2.Paf)}+\left\langle
\mathbf{D},\mathbf{E}^{(2.Paf)}\right\rangle _{L}\mathbf{D}_{s}
\label{12512} \\
&&-\left\langle \mathbf{D}_{s},\mathbf{E}^{(2.Paf)}\right\rangle _{L}\mathbf{%
D}.  \nonumber
\end{eqnarray}%
It follows that if the electric field $\mathbf{E}^{(2.Paf)}$ of the type-2 PAF is Fermi-Walker parallel  for third case, then      Eqs.(\ref{1010}) and (\ref{12512}) imply
\begin{equation}
\mathbf{E}_{s}^{(2.Paf)}=\left\langle \mathbf{D}_{s},\mathbf{E}%
^{(2.Paf)}\right\rangle _{L}\mathbf{D.}  \label{686812}
\end{equation}%
Also, the electric field $\mathbf{E}^{(2.Paf)}$ for third case with respect to the
type-2 PAF becomes
\begin{equation}
\mathbf{E}^{(2.Paf)}=\mathbf{E}^{(2.Paf)H}\mathbf{H}+\mathbf{E}^{(2.Paf)N}%
\mathbf{N},  \label{789*12}
\end{equation}%
where
\[
E^{(2.Paf)H}=\left\langle \mathbf{E}^{(2.Paf)},\mathbf{H}\right\rangle _{L}, \quad
E^{(2.Paf)N}=\left\langle \mathbf{E}^{(2.Paf)},\mathbf{N}%
\right\rangle _{L}.
\]%
Taking derivative of Eq.(\ref{789*12}) for $s$, the variation
of the electric field vector $\mathbf{E}^{(2.Paf)}$ is given by%
\begin{eqnarray}
\mathbf{E}_{s}^{(2.Paf)} &\mathfrak{=}&\mathbf{(}%
E_{s}^{(2.Paf)H}-p_{1}E^{(2.Paf)N})\mathbf{H}  \label{6751*1} \\
&&+\mathbf{(}p_{1}E^{(2.Paf)H}+E_{s}^{(2.Paf)N})\mathbf{N}  \nonumber \\
&&-\mathbf{(}p_{2}E^{(2.Paf)H}+p_{3}E^{(2.Paf)N})\mathbf{D}.  \nonumber
\end{eqnarray}%
From this,   
Eq. (\ref{6751*1}) implies
\begin{equation}
E_{s}^{(2.Paf)H}=p_{1}E^{(2.Paf)N},\quad %
E_{s}^{(2.Paf)N}=-p_{1}E^{(2.Paf)H},   \label{4566**}
\end{equation}%
it follows that Eq.(\ref{4566**}) gives  a rotation of the polarization plane by an angle $%
p_{1}$ for the third case with the type-2 PAF  in $\mathbb{R}_{1}^{3}.$

The Lorentz force equation $\Phi ^D\mathbf{E}^{(2.Paf)}$ of the electric
field vector $\mathbf{E}^{(2.Paf)}$ with the magnetic vector field $\mathbf{V%
}^{(3)}$\ is

\begin{equation}
\Phi ^{D}\mathbf{E}^{(2.Paf)}=\mathbf{E}_{s}^{(2.Paf)}=\mathbf{V}%
^{(3)}\times _{L}\mathbf{E}^{(2.Paf)}.  \label{123*69}
\end{equation}%
The curve $E^{(2.Paf)}V^{(3)}$ travelled along the electromagnetic
trajectory is called the electromagnetic curve with respect to
the type-2 PAF  in $\mathbb{R}_{1}^{3}$. The  Lorentz force
equations $\Phi ^D$ of the type-2 PAF for the optical fiber via Eqs.(\ref{345*78}) and (%
\ref{123*69}) for the third case  are
given by

\begin{equation}
\left[
\begin{array}{c}
\Phi ^{D}(\mathbf{H)} \\
\Phi ^{D}(\mathbf{N}) \\
\Phi ^{D}(\mathbf{D})%
\end{array}%
\right] =\left[
\begin{array}{ccc}
0 & -\xi & -p_{2} \\
\xi & 0 & -p_{3} \\
-p_{2} & -p_{3} & 0%
\end{array}%
\right] \left[
\begin{array}{c}
\mathbf{H} \\
\mathbf{N} \\
\mathbf{D}%
\end{array}%
\right], \label{7975**}
\end{equation}%
which implies that the magnetic vector field $\mathbf{V}^{(3)}$ via Eq.(\ref{7975**}) for third
case with respect to the type-2 PAF is expressed  by
\[
\mathbf{V}^{(3)}=-\xi \mathbf{D-}p_{2}\mathbf{N}+p\mathbf{_{3}H}.
\]


 \section{ The evolution of electric field for the type-3 PAF}

In this section, we consider   an optical fiber $\mathcal{O}$  described by the type-3 PAF  of  the timelike curve $\beta $
 in  Minkowski 3-space $\mathbb{R}_{1}^{3}$.

Suppose that the electric field vector $\mathbf{E}^{(3.Paf)}$ is
perpendicular to the timelike vector $\mathbf{B}$ according to the type-
2 PAF  in $\mathbb{R}_{1}^{3},$ that is,
\begin{equation}
\left\langle \mathbf{E}^{(3.Paf)}\mathbf{,B}\right\rangle _{L}=0.
\label{100}
\end{equation}%
Consider that no various loss mechanism along with the optical fiber for the
type-3 PAF.  Then  we have
\begin{equation}
\left\langle \mathbf{E}^{(3.Paf)}\mathbf{,E}^{(3.Paf)}\right\rangle _{L}=0.
\label{101}
\end{equation}%
The general variation of the electric field vector $\mathbf{E}^{(3.Paf)}$ for
the type-3 PAF is given by
\begin{equation}
\mathbf{E}_{s}^{(3.Paf)}=d_{1}\mathbf{P}+d_{2}\mathbf{F}+d_{3}\mathbf{B},
\label{102}
\end{equation}%
where $d_{1},$ $d_{2}$ and $d_{3}$ are the arbitrary smooth
functions. Via Eqs.(\ref{100}), (\ref{101}) and (\ref{102}), it is obtained by
\begin{eqnarray}
d_{1} &=&\lambda \left\langle \mathbf{E}^{(3.Paf)},\mathbf{F}\right\rangle_{L},\quad
d_{2}=-\lambda \left\langle \mathbf{E}^{(3.Paf)},\mathbf{P}%
\right\rangle _{L},  \label{103} \\
d_{3} &=&n_{2}\left\langle \mathbf{E}^{(3.Paf)},\mathbf{P}\right\rangle
_{L}-n_{3}\left\langle \mathbf{E}^{(3.Paf)},\mathbf{F}\right\rangle _{L},
\label{104}
\end{eqnarray}%
where $\left\langle \mathbf{E}^{(3.Paf)},\mathbf{P}\right\rangle _{L}\neq 0,$
\ $\left\langle \mathbf{E}^{(3.Paf)},\mathbf{F}\right\rangle _{L}\neq 0$ and
$\lambda $ is a parameter. If Eqs.(\ref{103}) and(\ref{104}) are substituted in
Eq.(\ref{102}), the change of the electric field vector $\mathbf{E}%
^{(3.Paf)} $ for the type-3 PAF  is found as
the following:
\begin{eqnarray}
\mathbf{E}_{s}^{(3.Paf)} &=&\lambda \left\langle \mathbf{E}^{(3.Paf)},%
\mathbf{F}\right\rangle _{L}\mathbf{P}-\lambda \left\langle \mathbf{E}%
^{(3.Paf)},\mathbf{P}\right\rangle _{L}\mathbf{F}  \label{105} \\
&&+\left(n_{3}\left\langle \mathbf{E}^{(3.Paf)},\mathbf{F}\right\rangle
_{L}-n_{2}\left\langle \mathbf{E}^{(3.Paf)},\mathbf{P}\right\rangle _{L}\right)%
\mathbf{B}.  \nonumber
\end{eqnarray}%
On the other hand, the Fermi-Walker derivative of the electric field $\mathbf{E}%
^{(3.Paf)}$ for the type-3 PAF  in $\mathbb{R}_{1}^{3}$ is expressed by
\begin{eqnarray}
^{FW}\mathbf{E}_{s}^{(3.Paf)} &=&\mathbf{E}_{s}^{(3.Paf)}-\left\langle
\mathbf{B},\mathbf{E}^{(3.Paf)}\right\rangle _{L}\mathbf{B}_{s}  \label{106}
\\
&&+\left\langle \mathbf{B}_{s},\mathbf{E}^{(3.Paf)}\right\rangle _{L}\mathbf{%
B}.  \nonumber
\end{eqnarray}%
If the electric field vector $\mathbf{E}^{(3.Paf)}$ of the  type-3 PAF is
Fermi-Walker parallel,
then Eqs.(\ref {100}) and (\ref{106}) lead to
\begin{equation}
\mathbf{E}_{s}^{(3.Paf)}=-\left\langle \mathbf{B}_{s},\mathbf{E}%
^{(3.Paf)}\right\rangle _{L}\mathbf{B.}  \label{107}
\end{equation}%
Also, the electric field vector $\mathbf{E}^{(3.Paf)}$ of the type-3 PAF is given by
\begin{equation}
\mathbf{E}^{(3.Paf)}=-\mathbf{E}^{(3.Paf)P}\mathbf{P}+\mathbf{E}^{(3.Paf)F}%
\mathbf{F},  \label{108}
\end{equation}%
where
\[
E^{(3.Paf)P}=\left\langle \mathbf{E}^{(3.Paf)},\mathbf{P}\right\rangle _{L}, \quad %
E^{(3.Paf)F}=\left\langle \mathbf{E}^{(3.Paf)},\mathbf{F}%
\right\rangle _{L}
\]%
it follows that
the evolution of the electric field vector
$\mathbf{E}^{(3.Paf)}$ is given by
\begin{eqnarray}
\mathbf{E}_{s}^{(3.Paf)} &\mathfrak{=}&\mathbf{(-}%
E_{s}^{(3.Paf)P}+n_{1}E^{(3.Paf)F})\mathbf{P}  \label{109} \\
&&+\mathbf{(-}n_{1}E^{(3.Paf)P}+E_{s}^{(3.Paf)F})\mathbf{F}  \nonumber \\
&&+\mathbf{(-}n_{2}E^{(3.Paf)P}+n_{3}E^{(3.Paf)F})\mathbf{B}.  \nonumber
\end{eqnarray}%
Furthermore, from Eqs.(\ref{107}) and (\ref{109}) we obtain
\begin{equation}
E_{s}^{(3.Paf)P}=n_{1}E^{(3.Paf)F},\quad %
E_{s}^{(3.Paf)F}=n_{1}E^{(3.Paf)P}, \label{110}
\end{equation}%
it gives a rotation of the polarization plane by an angle $n_{1}$
for the type-3 PAF  in $\mathbb{R}_{1}^{3}.$

The Lorentz force equation $\Phi ^B\mathbf{E}^{(3.Paf)}$ of the electric
field vector $\mathbf{E}^{(3.Paf)}$ with the magnetic vector field $\mathbf{W%
}$\ is

\begin{equation}
\Phi ^{B}\mathbf{E}^{(3.Paf)}=\mathbf{E}_{s}^{(3.Paf)}=\mathbf{W}\times _{L}%
\mathbf{E}^{(3.Paf)}.  \label{111}
\end{equation}%
The curve $E^{(3.Paf)}W$ moved along the electromagnetic trajectory is
called the  electromagnetic curve of  the type-3 PAF  in $%
\mathbb{R}_{1}^{3}$. The Lorentz force equations $\Phi ^B$ along the
optical fiber via Eqs.(\ref{105}) and (\ref{111}) for the type-3 PAF
are given by
\begin{equation}
\left[
\begin{array}{c}
\Phi ^{B}(\mathbf{P)} \\
\Phi ^{B}(\mathbf{F}) \\
\Phi ^{B}(\mathbf{B})%
\end{array}%
\right] =\left[
\begin{array}{ccc}
0 & \lambda & -n_{2} \\
\lambda & 0 & -n_{3} \\
-n_{2} & n_{3} & 0%
\end{array}%
\right] \left[
\begin{array}{c}
\mathbf{P} \\
\mathbf{F} \\
\mathbf{B}%
\end{array}%
\right].  \label{112}
\end{equation}%
The magnetic vector field $\mathbf{W}$ via Eq.(\ref{112}) for the type-3 PAF
is found as the following
\[
\mathbf{W}=-n\mathbf{_{3}P+}n_{2}\mathbf{F+}\lambda \mathbf{B}.
\]

\section*{Conclusion}

First of all, we construct the type-2 and type-3 PAFs in
Minkowski 3-space. Later, the evolutions of the electric field were
presented with respect to the the type-2 and the type-3 PAFs in the
Minkowski 3-space. Also the type-2 and the type-3 Lorentz equations, the
type-2 and the type-3 PAFs electromagnetic curves were given for the type-2
and the type-3 PAFs in the Minkowski 3-space

\section*{Acknowledgements.} N.E.G\"{u}rb\"{u}z was supported by the Scientific
Research Agency of Eski\c{s}ehir Osmangazi University (ESOGU BAP Project
Number: 202019016), and D.W. Yoon   was supported by the National Research Foundation of Korea (NRF) grant funded by the Korea government (MSIT) (No. 2021R1A2C101043211).

\section*{Declaration of Competing Interest}

The authors report no declarations of interest.

\section*{ Data Availability Statement}
Data sharing not applicable to this article as no datasets were generated or
analysed during the current study.

\end{document}